# Compressão de dados sem perdas para dispositivos IoT

Abraão Caldas[1], Renato Degelo[1], Edjair Mota[2], Celso B. Carvalho[1]

[1]Programa de pós-graduação em engenharia elétrica (PPGEE) – Universidade Federal do Amazonas (UFAM)
69080-900 – Manaus – AM – Brasil

[2]Instituto de Computação – Universidade Federal do Amazonas (UFAM)
69080-900 – Manaus – AM – Brazil

{abraaocaldas@gmail.com, rdegelo}@gmail.com, edjair@icomp.ufam.edu.br, ccarvalho_@ufam.edu.br

***Abstract.*** *In environments with energy and processing constraints, such as sensor networks and embedded devices, sending raw information over wireless networks can be costly. In order to reduce the amount of transmitted data and ultimately save energy, we can compress data before transmission. In this paper, we tackle such problem in the IoT domain by deploying two widely used libraries to deliver asynchronous messages and data compression/decompression. We evaluate both our methodology and the compression/decompression performance of the embedded library in a micro-controller for IoT.*

***Resumo.*** *Em ambientes com restrições de energia e processamento como as redes de sensores e dispositivos embarcados, o simples envio de informação através da rede sem fio pode ser muito custoso. Com o objetivo de diminuir a quantidade de dados transmitida e, consequentemente, economizar energia podemos comprimir os dados antes da transmissão. Neste artigo, embarcamos em dispositivo microcontrolador para IoT duas bibliotecas amplamente empregas atualmente para a entrega de mensagens assíncronas e a compressão/descompressão de dados. No trabalho também avaliamos o desempenho de compressão/descompressão da biblioteca embarcada.*

## 1. Introdução

Uma Rede de Sensores Sem Fio (RSSF) é composta por dispositivos sensores espalhados em um determinado local e que monitoram grandezas físicas ou eventos. Exemplos de grandezas físicas monitoradas são temperatura e humidade e exemplos de eventos são a movimentação de objetos alvo (TAVARES, 2015). Nas aplicações de RSSF os nós sensores possuem restrições de energia, processamento e armazenamento. Neste cenário, é fundamental o estudo e utilização de técnicas que possam reduzir o consumo de energia e a utilização de algoritmos e funcionalidades que possam ser embarcados nos dispositivos sensores, apesar de suas restrições. Como a maior parte do consumo de um dispositivo que se comunica utilizando o meio sem fio está na transmissão de dados e não no processamento de informação, torna-se vantajoso em termos de energia a compressão de dados antes da transmissão. Visando a redução do consumo de energia em dispositivos microcontroladores Raspberry Pi3 (MAKSIMOVIĆ, 2014) e que são utilizados em cenários de IoT (*Internet of Things*), nesta pesquisa, embarcamos nestes dispositivos, duas bibliotecas que são amplamente utilizadas em cenários de sistemas sem restrição de energia. A primeira biblioteca é baseada no protocolo ZeroMQ (SOMECH, 2017), chamada de NetMQ (NETMQ, 2017), ela tem como função a entrega de mensagens assíncronas entre dispositivos de sistemas distribuídos (HINTJENS, 2013). A segunda biblioteca denominada Snappy (GOOGLE, 2017) é atualmente empregada na compressão/descompressão de dados em sistemas cliente-servidor, ou seja, sem restrições de energia e/ou processamento.

Neste trabalho verificamos a viabilidade de embarcar as duas bibliotecas comentadas nos dispositivos Raspberry Pi 3. No artigo também avaliamos o desempenho de compressão de dados da biblioteca Snappy que embarcamos no dispositivo.

Para tratar os assuntos comentados, dividimos este artigo nas seguintes seções. A Seção 2 comenta sobre o consumo de energia em redes sem fio; A Seção 3 comenta sobre como funciona a compressão de dados utilizando a biblioteca Snappy; A Seção 4 cita as funcionalidades da biblioteca NetMQ; A Seção 5 apresenta o cenário de experimentação, os dispositivos utilizados e resultados de compressão de dados utilizando o microcontrolador Raspberry Pi3 e a biblioteca Snappy que embarcamos no dispositivo; Por fim, na Seção 6 apresentamos as conclusões e trabalhos futuros.

## 2. Consumo de energia em transmissões sem fio

Segundo Kimura (2005, p8-13), aproximadamente 80% da energia utilizado por um nó sensor é utilizada na transmissão de dados e, se a quantidade de dados transmitidos for minimizada, então o consumo de energia também diminuirá, mesmo que isso se traduza em um maior gasto de energia do processador.

Contudo isso nem sempre é verdade se o tempo de acesso a memória e o processamento for muito grande, ou seja, se o conjunto de instruções for lenta ou a compressão não reduzir significativamente o tamanho dos dados.

Neste artigo não consideramos perturbações de transmissão em redes sem fio para avaliar o desempenho de compressão da biblioteca Snappy. Desta forma, executamos a compressão de dados e realizamos a transmissão destes dados, sem perdas, em uma rede cabeada com o objetivo de avaliar a viabilidade de serem embarcados no microcontrolador Raspberry Pi as bibliotecas NetMQ e Snappy. Também avaliamos o desempenho de compressão da biblioteca Snappy desacoplando o impacto do meio de comunicação sem fio e suas perdas.

## 3. Compressão de dados usando Snappy

Snappy (GOOGLE, 2017) é uma biblioteca de compressão criada pelo Google. Seu objetivo não é ter a máxima compressão ou compatibilidade com outras bibliotecas e sim, objetiva alta velocidade com compressão mediana. Comparada com o melhor desempenho da biblioteca zLib, a biblioteca Snappy é muito mais rápida, com resultados entre 20% a 100% mais veloz (GOOGLE, 2017).

As principais características são: 1) Rapidez: Taxa de compressão igual ou superior à 250Mb/s; Estabilidade: Ao longo dos anos e sem grandes mudanças Snappy é usada para comprimir e descomprimir peta bytes no ambiente de produção da Google; 3) Segurança: Segura o suficiente para ser imune a ataques de corrupção de dados; 4) Segundo Kamburugamuve (2015, p7), em comparação com métodos tradicionais como LZ4 e JZLib, o Snappy foi mais rápido além de obter taxa de compressão de 10:1 em um tempo de $10ms$.

Para que a biblioteca seja utilizada no Windows 10 IoT é necessário apenas que o código fonte seja importado em um novo projeto do tipo UWP (*Universal Windows Platform*) sem que seja necessária nenhuma modificação no código fonte, necessitando apenas a compilação para a plataforma ARM.

## 4. NetMQ

NetMQ (SOMECH, 2017) é uma implementação em C# da biblioteca assíncrona de transferência de mensagens ZeroMQ (SOMECH, 2017). NetMQ foi escolhida para este artigo seguindo a avaliação de Dworak (2014) por motivos como: 1) API amigável e intuitiva; 2) Baixo custo de memória e alto desempenho; 3) Padrão de comunicação mais rápida que o TCP/Sockets puro; 4) Alto paralelismo de mensagens.

Com o objetivo de ser utilizada por outros pesquisadores, disponibilizamos (CALDAS, 2017) o código da biblioteca NetMQ que modificamos, portamos e embarcamos no microcontrolador Raspberry Pi3.

## 5. Experimento

Para avaliação da comunicação e compressão de dados, utilizamos como nó sensor uma Raspberry Pi3 equipada com sistema operacional Windows 10 IoT. A base de dados a ser lida e comprimida pelo nó sensor é a REDD (KOLTER, 2011) que possui 3.990.015 tuplas que representam leituras de valores de consumo de energia elétrica de aparelhos domésticos de uma residência. Cada leitura ou tupla da base REED é composta por informações de instante de tempo da leitura (8 bytes) e consumo de energia do dispositivo em formato *double*. Utilizamos como *sink* (canalizador) da rede de sensores um computador core i7-Skylake com 16GB RAM equipado com Windows 10. O sensor e o *sink* se comunicam através de conexão cabeada com velocidade de 100Mbps.

### 5.1 Arquitetura

Conforme figura 1, contabilizamos 03 diferentes tempos para a avaliação de desempenho. O primeiro tempo medido, contabiliza o tempo de compressão mais o tempo de transmissão dos dados, o segundo tempo medido contabiliza somente o tempo de transmissão dos dados que foram comprimidos em etapa anterior e o terceiro tempo medido, contabiliza somente o tempo de transmissão dos dados não comprimidos.

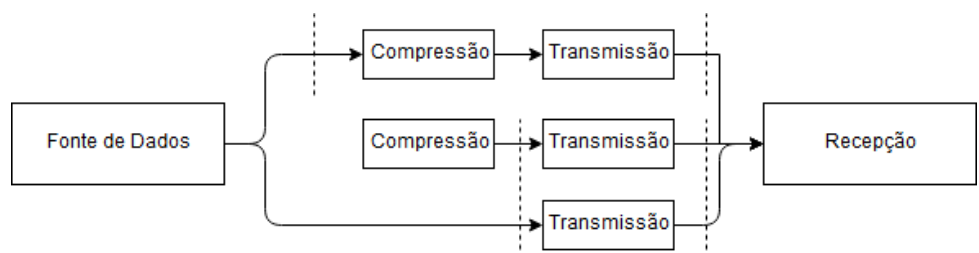

**Figura 1. Diagrama demonstrando o fluxo de dados e tempos medidos na avaliação de desempenho.**

## 5.2 Resultados

Na Figura 2, no eixo X temos a quantidade de tuplas de dados lidas de uma única vez (exemplo: 1000, 200, 100 e 20) da base dados REDD. No eixo Y, apresenta-se o tempo, utilizado no processamento e/ou transmissão dos dados. Na mesma figura a curva cinza representa o tempo gasto com a compressão mais transmissão de dados comprimidos. A curva laranja o tempo de transmissão dos dados que foram pré-comprimidos e a curva azul o tempo de transmissão dos dados sem compressão.

Na maioria dos casos o tempo de transmissão dos dados pré-comprimidos (curva laranja) foi menor que a transmissão sem compressão (curva azul), no entanto fazendo uso excessivo de CPU conforme notado na curva cinza. O único cenário em que a compressão mais transmissão (curva cinza) apresenta tempo próximo do cenário de transmissão sem compressão utiliza 1000 tuplas/leituras. Para valores menores que este, a compressão não é viável. No entanto, comenta-se que em cenários de IoT de comunicações sem fio, o gasto de energia com transmissão é muito superior ao gasto de energia com processamento (RAZZAQUE, 2014), sendo necessário avaliar futuramente o consumo de energia com e sem compressão.

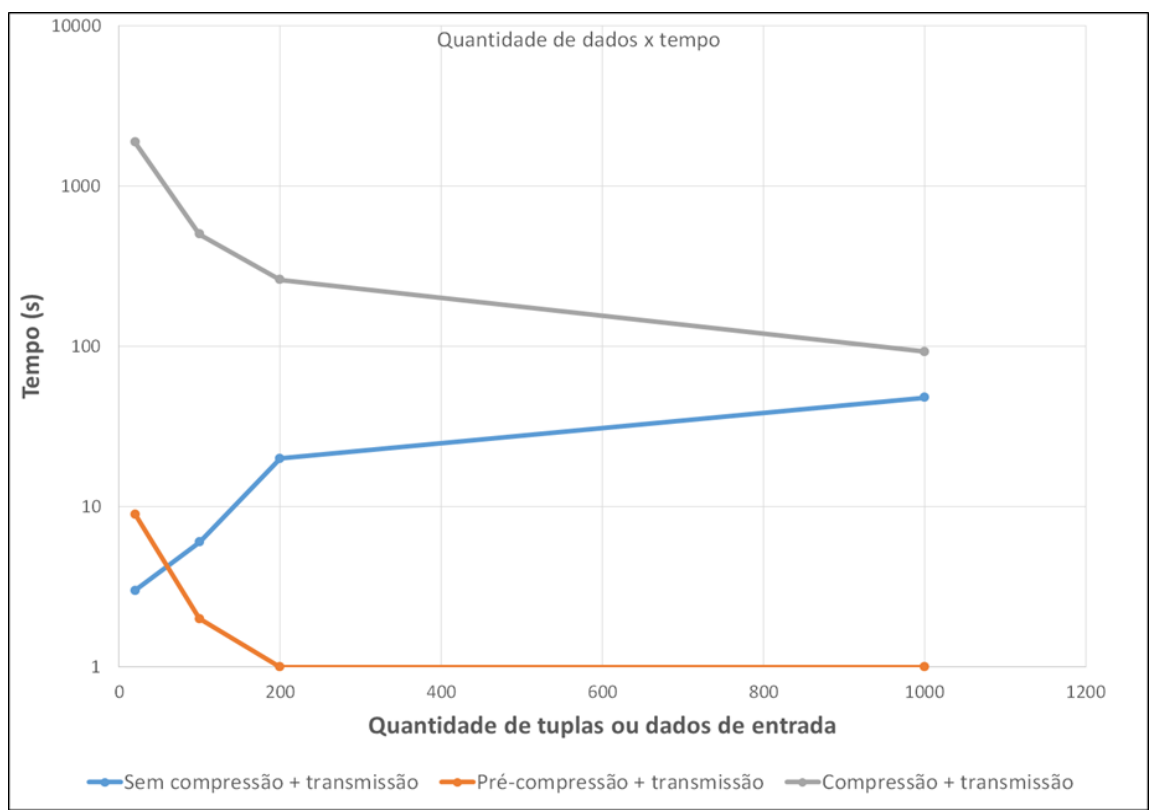

**Figura 2. Quantidade de dados transmitidos versus tempo (em segundos).**

Tabela 1. Dados coletados em segundos.

| Quantidade de dados | Compressão + Transmissão | Pré-Compressão + Transmissão | Sem Compressão + Transmissão |
|---|---|---|---|
| 1000 | 93 | 1 | 48 |
| 200 | 261 | 1 | 20 |
| 100 | 501 | 2 | 6 |
| 20 | 1878 | 9 | 3 |

Como podemos observar a medida que aumentamos a quantidade de dados transmitidos de uma só vez também nos aproximamos do mesmo tempo de transmissão dos mesmos dados sem compressão, se levarmos em conta os dados pré-comprimidos, podemos observar que o tempo de transmissão de 1000 tuplas sem compressão é de 48 segundos e o tempo de tranmissão de dados comprimidos (sem levar em consideração o tempo de compressão) é de apenas 1 segundo.

Segundo Kimura (2005, p8-13), a transmissão de um bit de informação é 480 vezes mais custosa eletricamente que o processamento de um instrução de dados, então podemos dizer que neste caso comprimir dados foi eletricamente eficiente.

Apesar do resultado positivo podemos verificar que este resultado pode ser melhorado levando em consideração que:

- O Snappy usa instruções 64-bits (GOOGLE, 2017) , em ambientes 32 bits é necessário emular essas instruções;
- Neste trabalho portamos a biblioteca Snappy para UWP (Universal Windows Platform), por esse motivo o Snappy pode não estar otimizado para rodar em processador ARM;
- Não utilizamos processamento paralelo, a mesma *thread* que compacta os dados os transmite.

## 6. Conclusões e trabalhos futuros

Neste artigo embarcamos em um microcontrolador Raspberry Pi3 duas bibliotecas, comumente utilizadas em dispositivos sem restrições de memória, armazenamento ou energia. A primeira biblioteca chamada NetMQ executa a entrega de mensagens assíncronas e a segunda chamada Snappy executa compressão/descompressão de dados. Verificamos a viabilidade de embarcar as bibliotecas no dispositivo IoT Raspberry Pi3. Ainda existe um grande campo para otimização de algoritmos de compressão em dispositivos IoT, neste caso o Snappy ainda carece de um ajuste fino para rodar em um ambiente tão limitado. Também é necessário comparar o desempenho de compressão do algoritmo Snappy com outros algoritmos de compressão. Outros estudos futuros, visam otimizar as instruções para executar o Snappy em um microcontrolador ARM como o da Raspberry pi3. Além disso, o estudo do consumo de energia do processamento de compressão de dados utilizando o Snappy em ambientes de comunicação sem fio, fica em aberto.

Em trabalhos futuros é interessante também a avaliação comparativa com outros algoritmos de compressão como os algoritmos da família PAQ, LZO e BZip.

# 7. Referências bibliográficas